\documentclass[final,5p,times,twocolumn]{elsarticle}
\biboptions{sort&compress}

\usepackage{amssymb}
\usepackage{lipsum}
\usepackage{amsmath}
\usepackage{hyperref}
\setcitestyle{square}

\journal{Nuclear Instruments and Methods in Physics Research Section A}

\usepackage{fancyhdr}

\begin{document}
\frenchspacing
\begin{frontmatter}



\title{Optical Characterization Measurements of Commercial Polyvinyltoluene Scintillators for Accurate Geant4 Detector Modeling}


\author[first,second]{Fergus M. Huggins}
\author[second]{Micah S. Johnson}
\author[second]{Matthew Q. Buckner}
\affiliation[first]{organization={Colorado School of Mines Department of Physics},
            addressline={1523 Illinois Street}, 
            city={Golden},
            postcode={80401}, 
            state={Colorado},
            country={USA}}
\affiliation[second]{organization={Lawrence Livermore National Laboratory},
            addressline={7000 East Avenue}, 
            city={Livermore},
            postcode={94550}, 
            state={California},
            country={USA}}
\begin{abstract}
Accurate Monte Carlo simulation of plastic scintillator detectors requires precise wavelength-dependent optical properties. We present measurements of wavelength-dependent refractive index $n(\lambda)$ and bulk absorption length $L_{\mathrm{abs}}(\lambda)$ for six commercial polyvinyltoluene (PVT)-based scintillators (Saint Gobain BC-400, BC-404, BC-408, BC-418, BC-422, and BC-430) at five wavelengths between 372 and 660~nm. Using a precision refraction apparatus and variable-thickness transmission measurements, we characterize material-specific dispersion that deviates substantially from both the manufacturer specification and literature values for pure PVT. Illustrative Geant4 simulations of a BC-422 detector demonstrate a 7.2\% reduction in photocathode energy-deposition threshold events when using our measured $n(\lambda)$ compared to the constant manufacturer value, highlighting the practical importance of accurate optical property data for detector design and calibration. We successfully measure absorption lengths for BC-430 at four wavelengths between 372 and 505 nm (2.2--12.4~mm) and for all materials at 372~nm (2.2--38~mm), while establishing lower bounds exceeding 750~mm for weakly absorbing formulations at longer wavelengths.
\end{abstract}




\end{frontmatter}




\section{Introduction}
\label{introduction}

Polyvinyltoluene (PVT)-based plastic scintillators are widely used in particle detection and radiation measurements \cite{Cushman2021plastic,HamelPlasticScintillators}. Their fast scintillation response, mechanical robustness, and cost-effective production make them useful for a range of detector applications \cite{HamelPlasticScintillators}. In typical detector architectures, the scintillator is optically coupled to a photosensor, such as a photomultiplier tube (PMT), silicon photomultiplier (SiPM), or photodiode. Detector response depends in part on the transport of scintillation photons from their production sites to the photosensitive region. 

Monte Carlo toolkits such as Geant4 \cite{geant4_paper} support detector design and performance studies by simulating particle interactions and optical-photon transport. The accuracy of these simulations depends on the material properties and optical interfaces used to describe the detector \cite{Khodaei2023geant4}. The wavelength-dependent refractive index, $ n(\lambda) $, governs refraction and reflection at material boundaries, while the bulk absorption length, $ L_{\mathrm{abs}}(\lambda) $, characterizes photon absorption during propagation. Their influence on the detected signal depends on the scintillation emission spectrum, detector geometry, surface treatment, optical coupling, and photosensor spectral response. Reliable optical-property inputs are therefore important for predicting light-collection efficiency and photon arrival-time distributions, although they are not the only sources of modeling uncertainty.

Manufacturer specifications provide useful nominal properties but may lack the wavelength-dependent data required for optical transport simulations. For example, the data sheet covering BC-400, BC-404, and BC-408 specifies a refractive index of $ n=1.58 $ without reporting its wavelength dependence \cite{bc40040408_datasheet}. A constant refractive index does not represent material dispersion, and its adequacy depends on the wavelengths and detector configuration being modeled.

Previous optical characterization studies include measurements of pure PVT \cite{nakamura2015optical} and temperature-dependent characterization of the commercial PVT scintillator BC-412 \cite{Cowles2022BC412}. Commercial formulations incorporate proprietary dopant concentrations and wavelength shifters that produce material-specific optical behavior \cite{HamelPlasticScintillators}. Consequently, measurements of the base polymer or a single commercial formulation do not, by themselves, establish the optical properties of other formulations. Additional formulation-specific measurements provide useful constraints for detector modeling.

In this work, we report refractive-index measurements and bulk absorption-length measurements or lower bounds for six commercial PVT scintillators: BC-400, BC-404, BC-408, BC-418, BC-422, and BC-430, listed in Table~\ref{tab:bc4xx_specs}. Measurements were performed at five laser wavelengths between 372 and 660~nm. Refractive indices were determined from lateral beam displacement through rotated samples using a method based on Ref.~\cite{Leite2021measurement}, while absorption lengths were inferred from variable-thickness transmission measurements. The apparatus and uncertainty treatment are described in Section~\ref{methodology}. We provide empirical refractive-index dispersion fits for use within the measured wavelength interval and present an illustrative Geant4 comparison of constant and wavelength-dependent refractive-index models. Scintillator bulk absorption is omitted from this comparison to assess sensitivity to the refractive-index input.

\begin{table}[ht]
  \centering
  \small
  \caption{Selected characteristics of the PVT scintillators studied
  in this work
  \cite{Cushman2021plastic,bc40040408_datasheet,bc418422_datasheet,bc430_datasheet}.
  Light-output values are manufacturer-reported percentages relative
  to anthracene.}
  \label{tab:bc4xx_specs}
  \setlength{\tabcolsep}{3pt}
  \begin{tabular*}{\columnwidth}{@{\extracolsep{\fill}}ccc@{}}
    \hline
    Material & Application or feature & Light output (\%) \\
    \hline
    BC-400 & General purpose      & 65 \\
    BC-404 & Fast counting        & 68 \\
    BC-408 & Time-of-flight       & 64 \\
    BC-418 & Ultra-fast timing    & 67 \\
    BC-422 & Very fast timing     & 55 \\
    BC-430 & Red-shifted emission & 45 \\
    \hline
  \end{tabular*}
\end{table}

\section{Methodology}
\label{methodology}
\subsection{Refraction Measurements}

\begin{figure}[ht]
  \centering
  \includegraphics[width=0.45\textwidth]{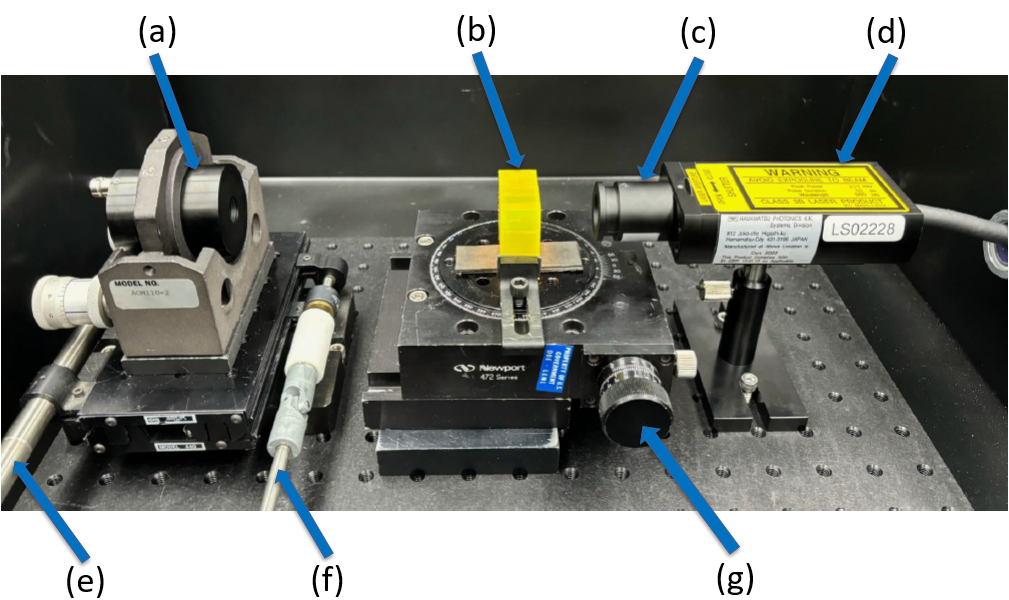} 
  \caption{Experimental apparatus used for refraction and attenuation measurements. (a) Hamamatsu MCP-PMT R3809U-51 PMT. (b) 40~mm $\times$ 40~mm $\times$ 20~mm BC-430 scintillator. (c) Hamamatsu A10089 aspheric Lens. (d) M10306-43 Hamamatsu laser head. (e) PMT translation stage. (f) PMT translation gauge (g) Newport 472 series rotation stage.}
  \label{fig:apparatus}
\end{figure}

We measure the wavelength-dependent refractive index using the refraction apparatus shown in Figure~\ref{fig:apparatus}. The system consists of a Hamamatsu R3809U-51 MCP-PMT with a Thorlabs S5K 5~$\mu$m slit, mounted on a precision translation stage with 0.01~mm resolution, and a Hamamatsu PLP-10 C10196 picosecond light pulser with interchangeable laser heads covering 372--660~nm as listed in Table~\ref{tab:wave}. Laser light passes through an aspheric lens, Hamamatsu A10089, to produce a collimated beam that enters the scintillator volume. The scintillator is mounted on a Newport 472-series rotation stage with angular resolution of $0.1^\circ$ that allows precise angular control. A Keysight DSOS404A oscilloscope records both PMT count rate and voltage waveforms throughout each measurement. Flat black absorbing paint was applied to the interior of the enclosure to prevent reflections and scattering of light for both quality data collection and safety.

\begin{figure}[ht]
  \centering
  \includegraphics[width=0.48\textwidth]{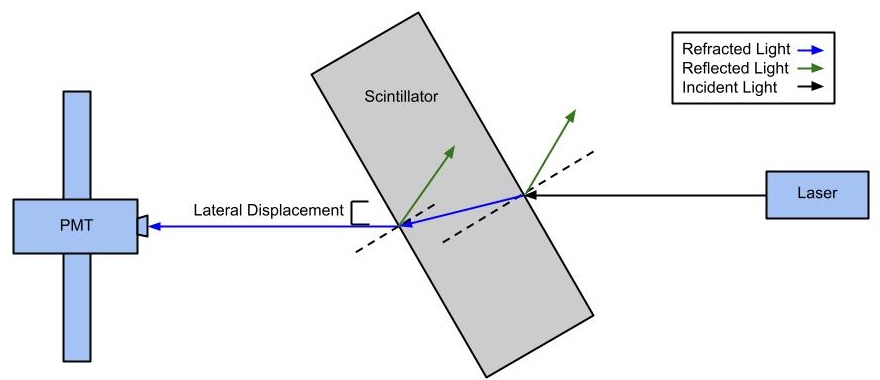} 
  \caption{Example diagram of light propagation through a scintillator rotated by 25 degrees. The lateral displacement of the refracted light is measured by the PMT, relative to normal incidence. Reflected light makes up a small portion of total intensity and is absorbed by the laser enclosure.}
  \label{fig:sketch}
\end{figure}

The measurement procedure begins with the scintillator positioned at normal incidence, $ 0^\circ $. We translate the PMT laterally to locate the beam edges, defined as the positions where the count rate is halfway between the illuminated plateau and the background level. We then rotate the scintillator in $5^\circ$ increments from $0^\circ$ to $+45^\circ$, recording the lateral beam displacement at each angle as shown in Figure~\ref{fig:sketch}. The procedure is repeated for negative angles from $0^\circ$ to $-45^\circ$. At each angle, we record the count rate profile across both edges of the refracted beam. We repeat this procedure for all six scintillator formulations listed in Table~\ref{tab:bc4xx_specs} at each of the five wavelengths in Table~\ref{tab:wave}. All measurements are performed in a light-tight enclosure to eliminate ambient light and ensure laser safety compliance.

\begin{table}[ht]
    \centering
    \caption{Laser heads used for refractive index measurements. Wavelengths have uncertainty of $\pm5$~nm, per Hamamatsu \cite{hamamatsu_plp10_datasheet}.}
    \label{tab:wave}
    \begin{tabular*}{\columnwidth}{@{\extracolsep{\fill}}ccc}
        \hline
         Model & Wavelength  & Nominal Color\\
        \hline
       M10306-27 & 372~nm  & Violet\\
       M10306-31 & 439~nm  & Blue\\
       M10306-35 & 482~nm  & Cyan\\
       M10306-37 & 505~nm  & Green\\
       M10306-43 & 660~nm  & Red\\
        \hline
    \end{tabular*}
\end{table}

\subsection{Refraction Analysis}

\begin{figure*}[ht]
  \centering
  \includegraphics[width=1.0\textwidth]{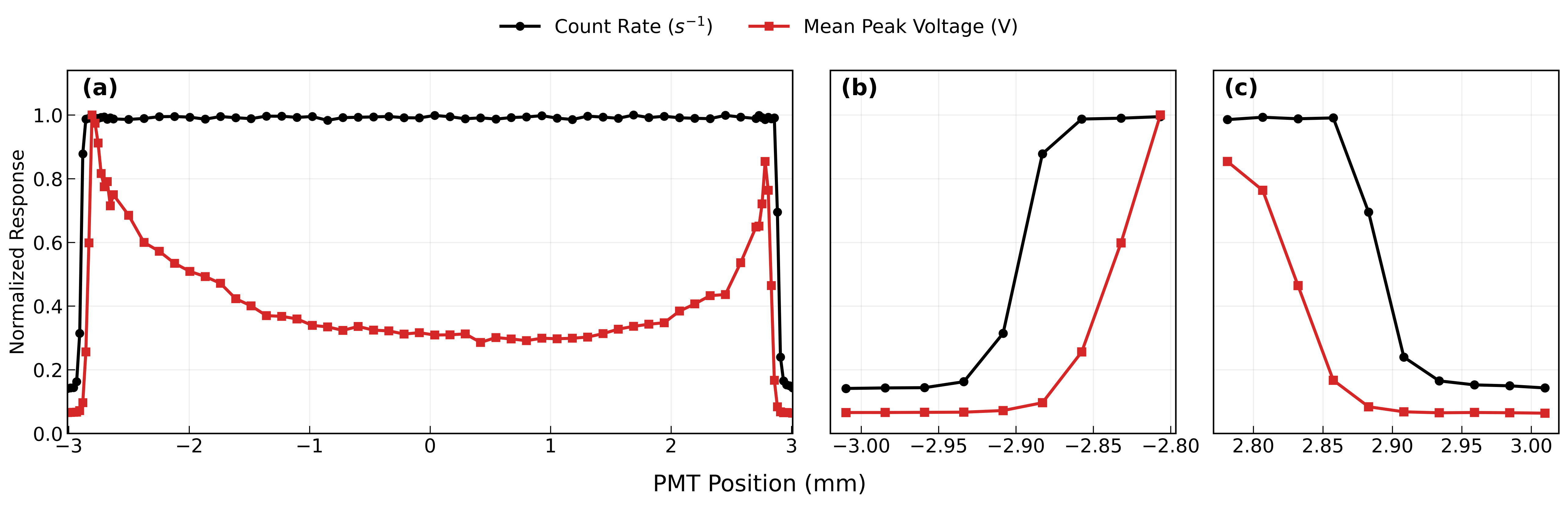} 
  \caption{(a) Count rate and mean peak voltage laser waveforms measured for a 40~mm $\times$ 40~mm $\times$ 40~mm BC-418 Scintillator at 482~nm and 100~Hz. PMT saturation causes drop in voltage response in the middle of the beam. (b) Left-edge profile, zoomed in. (c) Right-edge profile, zoomed in.}
  \label{fig:Edges}
\end{figure*} 

We determine the refractive index, $ n_s $, from the transverse beam displacement $ x $, the incidence angle $ \theta $, the scintillator thickness $ d $, and the refractive index of air $ n_a $ \cite{Singh2002refractive,Leite2021measurement}. Let $ p $ denote the lateral position of the PMT translation stage. For a given beam edge, $ p_{\mathrm{pre}} $ is its position with the scintillator at normal incidence, and $ p_{\mathrm{post}} $ is its position after rotation. The displacement is therefore $ x \equiv p_{\mathrm{post}}-p_{\mathrm{pre}} $, with the scintillator present in both measurements. For each measurement, we compute:
\begin{equation}
n_s = n_a \sqrt{
\frac{\cos^2\theta}
{\left(1-\frac{x}{d\sin\theta}\right)^2}
+\sin^2\theta
}.
\label{eqn:refraction}
\end{equation}

Approximating $n_a\approx1$ \cite{stone2001index}, uncertainties in $n_s$ arise from $\theta$, $d$, and $x$. We treat the angular and thickness uncertainties as purely systematic: $\sigma_{\theta} = 0.1^\circ$ is set by the rotation stage resolution, and $\sigma_d = 0.005$~mm follows the Saint-Gobain manufacturing tolerance \cite{SaintGobainData}. The lateral displacement $x$ has both systematic and statistical components. The systematic uncertainty $\sigma_{x,\mathrm{sys}} = 0.18$~mm combines the manufacturer-specified micrometer accuracy of 0.127~mm per measurement with the factor of $\sqrt{2}$ arising from the difference between pre-rotation and post-rotation edge positions. These systematic contributions propagate through the standard first-order error formula to yield the systematic component of $\sigma_{n_s}$. Improved angular and translation-stage metrology would reduce these instrumental uncertainty contributions in future measurements.

We estimate the statistical uncertainty in $ x $ from the count-rate measurements at the beam edges. The beam edge is defined as the position where the measured count rate reaches the midpoint between the illuminated plateau and background:
\begin{equation}
    C_{1/2} = C_{\mathrm{background}}
    + \frac{C_{\mathrm{plateau}} - C_{\mathrm{background}}}{2}.
    \label{eq:edge_midpoint}
\end{equation}
Individual measurements $ C_1 $ and $ C_2 $, obtained at positions $ p_1 $ and $ p_2 $, bracket the edge and are used
in the local linear approximation for estimating its statistical uncertainty. They do not define the plateau and background levels. We estimate the edge location and its variance via Fisher information \cite{Fisher1922}. The resulting statistical uncertainty is

\begin{equation}
\sigma^2_{p_{\mathrm{edge}}}
\approx
\frac{C_1 + C_2}{(\Delta C)^4}
\left[
\frac{(C_1 - C_{1/2})^2}{C_1}
+
\frac{(C_2 - C_{1/2})^2}{C_2}
\right]
(\Delta p)^2,
\end{equation}

where $\Delta C = C_2 - C_1$ and $\Delta p = p_2 - p_1$. This expression captures the Poisson scaling and the dependence on local contrast and spacing of the edge points. 

\begin{figure}[ht]
  \centering
  \includegraphics[width=0.45\textwidth]{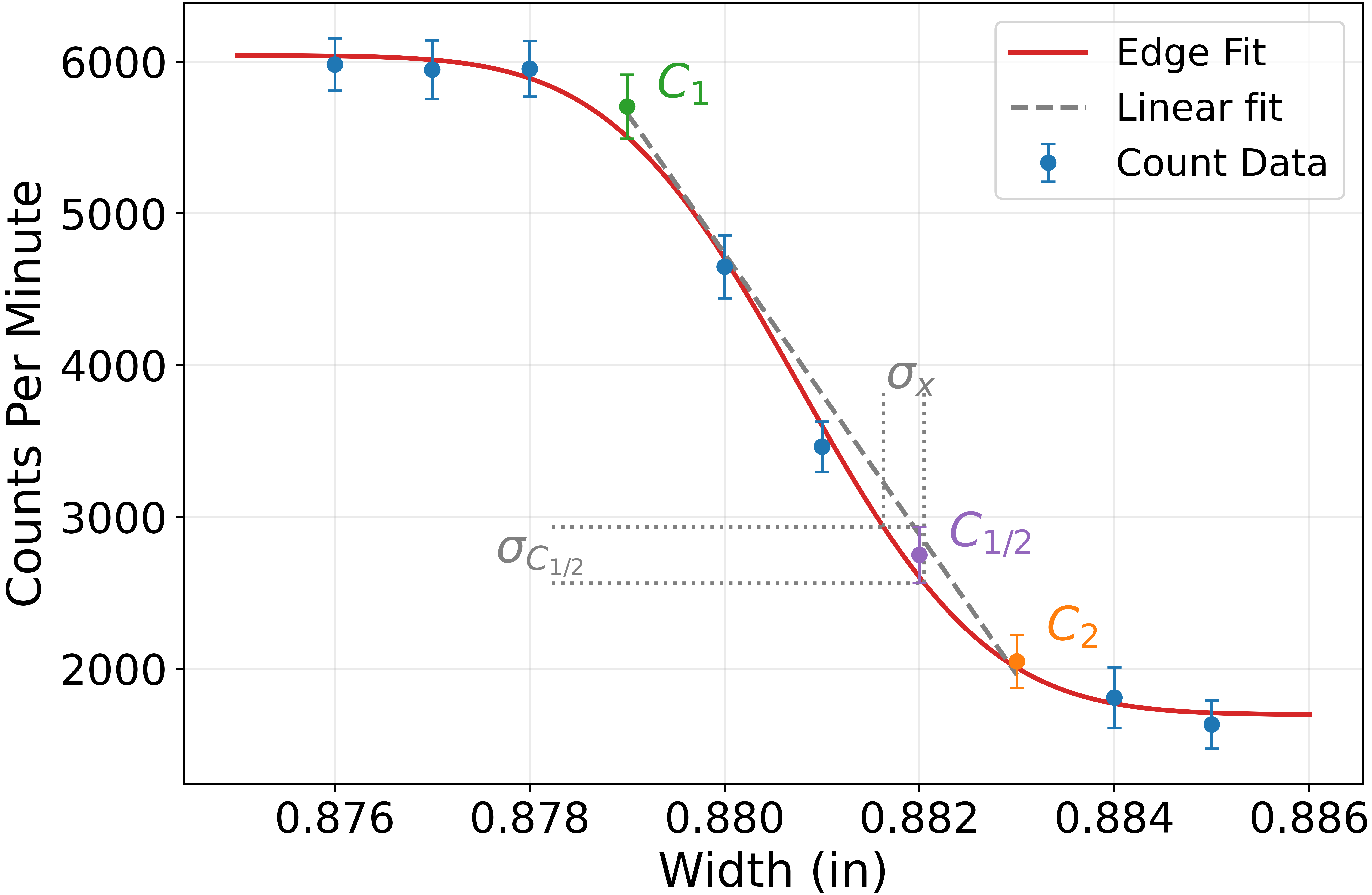} 
  \caption{Example beam-edge profile for a nominal 439~nm laser transmitted through a $ 40 \times 40 \times 40~\mathrm{mm}^3 $ BC-430 scintillator at normal incidence ($ \theta=0^\circ $). The measured count rates $ C_1 $, $ C_{1/2} $, and $ C_2 $ and their uncertainties are used to estimate the statistical uncertainty in the edge position.}
  \label{fig:edge}
\end{figure} 

We combine the systematic and statistical contributions to obtain the total uncertainty $\sigma_x = \sqrt{\sigma_{x,\mathrm{sys}}^2 + \sigma_{x,\mathrm{stat}}^2}$. We compute $n_s(\theta)$ and $\sigma_{n_s}(\theta)$ for each angle and wavelength. We find that small incidence angles yield large fractional uncertainties in $n_s$ because the beam displacement is comparable to the position and angle uncertainties. At larger angles, the displacement grows relative to its uncertainty and the resulting $n_s$ values carry smaller relative errors.

To obtain a single index of refraction for each scintillator and wavelength, we form a weighted average over angles using inverse-variance weighting,

\begin{equation}
\bar{n}_s =
\frac{\sum_i n_s(\theta_i)\,\sigma_{n_s}^{-2}(\theta_i)}
{\sum_i \sigma_{n_s}^{-2}(\theta_i)},
\end{equation}
with total uncertainty
\begin{equation}
\sigma_{\bar{n}_s}
=
\left(
\sum_i \sigma_{n_s}^{-2}(\theta_i)
\right)^{-1/2}.
\end{equation}

\begin{figure}[ht]
  \centering
  \includegraphics[width=0.48\textwidth]{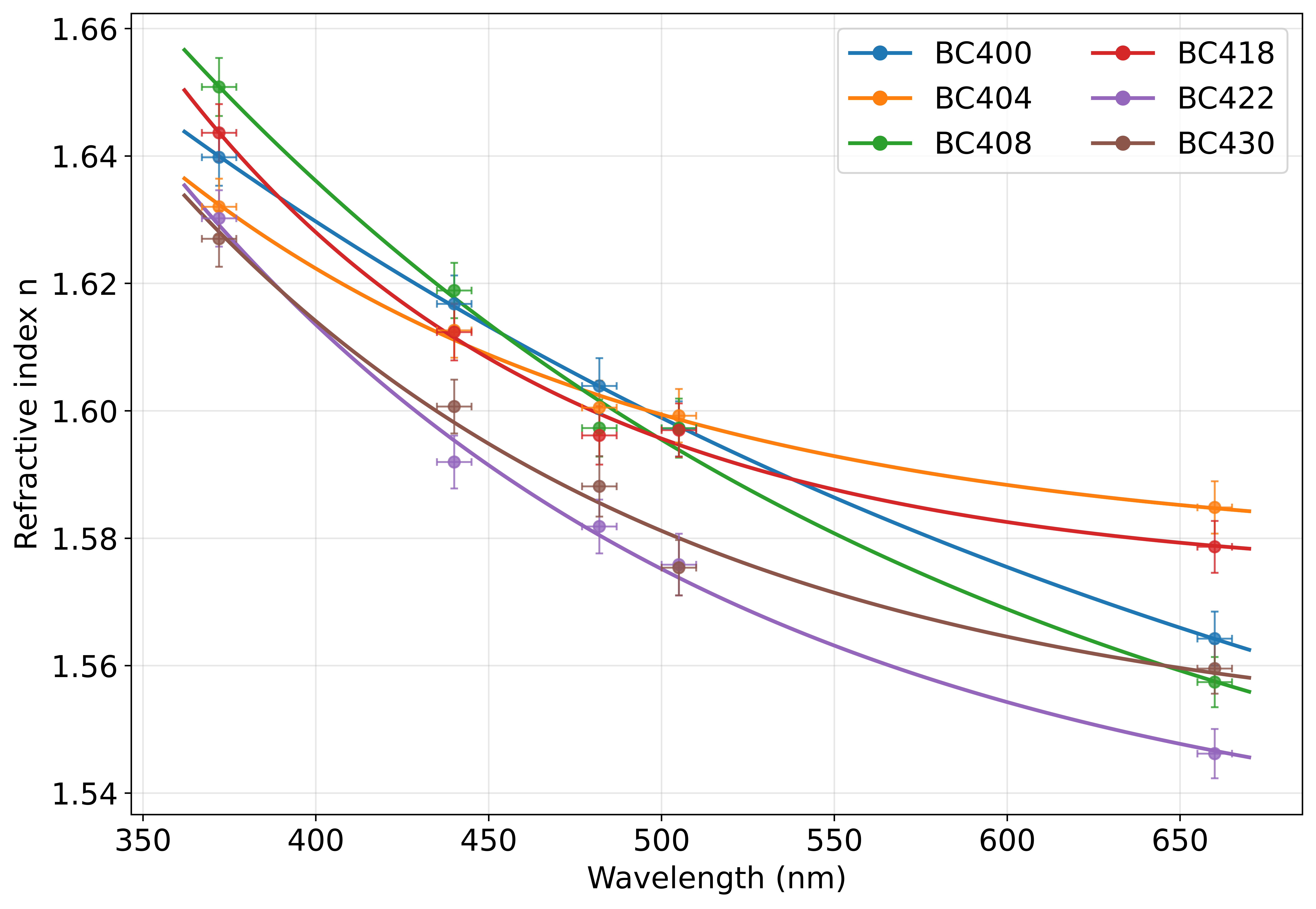} 
  \caption{Measured refractive index values and ODR curves for all six polyvinyltoluene formulations. Curves are defined over the 370-660~nm range. Coefficients for fit functions given in Table~\ref{tab:exp_fit_coeffs}. The formulations exhibit distinct wavelength-dependent refractive indices that are not represented by a single constant value.}
  \label{fig:DNL}
\end{figure}

This procedure naturally down-weights measurements at small incidence angles, where the smaller beam displacement results in a larger propagated uncertainty in the refractive index. Once we obtain the refractive index values and their uncertainties for each wavelength, we fit to an exponential model to describe the wavelength dependence, Equation~\ref{eqn:expntl}. We use an orthogonal distance regression (ODR) model \cite{boggs1990odr} to determine coefficients A, B, and C for each material. The measured refractive indices and ODR fits are shown in Figure~\ref{fig:DNL}.

\begin{equation}
n_{s,\lambda}=A+Be^{-C\lambda}
\label{eqn:expntl}
\end{equation}

We calculate ODR functions for the six formulations listed in Table~\ref{tab:bc4xx_specs}. We use ODR rather than ordinary least squares because both wavelength and refractive index carry uncertainties that should be accounted for in the fit. The coefficients are listed in Table~\ref{tab:exp_fit_coeffs}.

\begin{table}[ht]
\centering
\caption{Exponential fit coefficients for the six measured PVT formulations, using Equation~\ref{eqn:expntl} and ODR.}
\begin{tabular*}{\columnwidth}{@{\extracolsep{\fill}}cccc}
\hline
Material & $A$ & $B$ & C (nm$^{-1}$) \\
\hline
BC-400 & 1.501 $\pm$ 0.005 & 0.385 $\pm$ 0.009 & 0.00274 $\pm$ 0.00015\\
BC-404 & 1.578 $\pm$ 0.004 & 0.82 $\pm$ 0.31 & 0.0073 $\pm$ 0.0012\\
BC-408 & 1.518 $\pm$ 0.020 & 0.64 $\pm$ 0.20 & 0.0042 $\pm$ 0.0012 \\
BC-418 & 1.573 $\pm$ 0.005 & 2.03 $\pm$ 1.10 & 0.0090 $\pm$ 0.0015 \\
BC-422 & 1.529 $\pm$ 0.008 & 0.96 $\pm$ 0.32 & 0.0061 $\pm$ 0.0011\\
BC-430 & 1.548 $\pm$ 0.010 & 1.02 $\pm$ 0.68 & 0.0068 $\pm$ 0.0019\\
\hline
\end{tabular*}
\label{tab:exp_fit_coeffs}
\end{table}

\subsection{Attenuation Measurements}

For attenuation measurements, we use the apparatus described in Section~2.1 and shown in Figure~\ref{fig:apparatus}. We determine $ L_{\mathrm{abs}}(\lambda) $ using variable-thickness transmission measurements for the six formulations listed in Table~\ref{tab:bc4xx_specs}. At each wavelength, the integrated pulse area is measured for samples with thicknesses $ d_i $. The data $ \{d_i,T(\lambda,d_i)\} $ are fitted using Equation~\ref{eqn:bulkattn}, with $ T_0(\lambda) $ and $ L_{\mathrm{abs}}(\lambda) $ treated as fit parameters.

\begin{equation}
T(\lambda,d)=T_0(\lambda)e^{-d/L_{\mathrm{abs}}(\lambda)}
\label{eqn:bulkattn}
\end{equation}

We adjust the PMT bias voltage to -2.5~kV for attenuation measurements, compared to -3.3~kV for refraction measurements, to avoid saturation and keep the pulse area within the linear response range. The Keysight DSOS404A oscilloscope calculates the integrated pulse area for each pulse over a one minute period, along with its uncertainty. We repeat this measurement for material thicknesses of 1, 5, 10, 20, and 40~mm. Material thicknesses were obtained from the Saint-Gobain material data sheet \cite{SaintGobainData}. 

For some material and wavelength combinations, the change in transmitted pulse area over the available sample thicknesses is too small to determine a finite absorption length within the measurement uncertainty. In these cases, we report lower bounds using the adopted $ \Delta\chi^2=2.71 $ criterion. Absorption-length estimates and lower bounds are listed in Table~\ref{tab:absorptionlngths}.

\begin{table*}[htbp]
\centering
\tiny
\caption{Bulk absorption-length estimates and lower bounds. Finite estimates are reported with uncertainties; lower bounds use the $ \Delta\chi^2=2.71 $ criterion. Additional 100~mm samples were used for BC-400, BC-404, and BC-408.}
\label{tab:absorptionlngths}
\resizebox{\textwidth}{!}{%
\begin{tabular}{cccccc}
\hline
Material & 372 nm & 439 nm & 482 nm & 505 nm & 660 nm \\
\hline
BC-400 & $26.9 \pm 3.9$ mm & $>2.0\times10^3$ mm & $>1.6\times10^3$ mm & $>1.2\times10^3$ mm & $>2.2\times10^3$ mm  \\
BC-404 & $26.6 \pm 3.0$ mm & $>1.2\times10^3$ mm  & $>1.5\times10^3$ mm & $>1.3\times10^3$ mm  & $>9.7\times10^2$ mm  \\
BC-408 & $28.2 \pm 2.7$ mm & $>7.6\times10^2$ mm  & $>1.1\times10^3$ mm  & $>1.6\times10^3$ mm  & $>7.7\times10^2$ mm  \\
BC-418 & $21.2 \pm 3.4$ mm & $>7.9\times10^2$ mm  & $>1.0\times10^3$ mm  & $>8.8\times10^2$ mm  & $>8.7\times10^2$ mm  \\
BC-422 & $37.7 \pm 2.7$ mm & $>1.0\times10^3$ mm  & $>1.3\times10^3$ mm & $>8.8\times10^2$ mm  & $>9.3\times10^2$ mm  \\
BC-430 & $2.23 \pm 0.09$ mm & $9.03 \pm 0.15$ mm & $6.77 \pm 0.62$ mm & $12.42 \pm 0.20$ mm & $>8.3\times10^2$ mm \\
\hline
\end{tabular}%
}
\end{table*}

\section{Discussion}
\subsection{Refractive Index: Wavelength Dependence and Implications}

We observe substantial wavelength dependence in the refractive index across all six PVT-based scintillators. This demonstrates that the manufacturer-specified constant value of $n = 1.58$ represents at best a rough average and fails to capture the dispersive behavior essential for accurate optical transport modeling.

\begin{figure}[ht]
  \centering
  \includegraphics[width=0.48\textwidth]{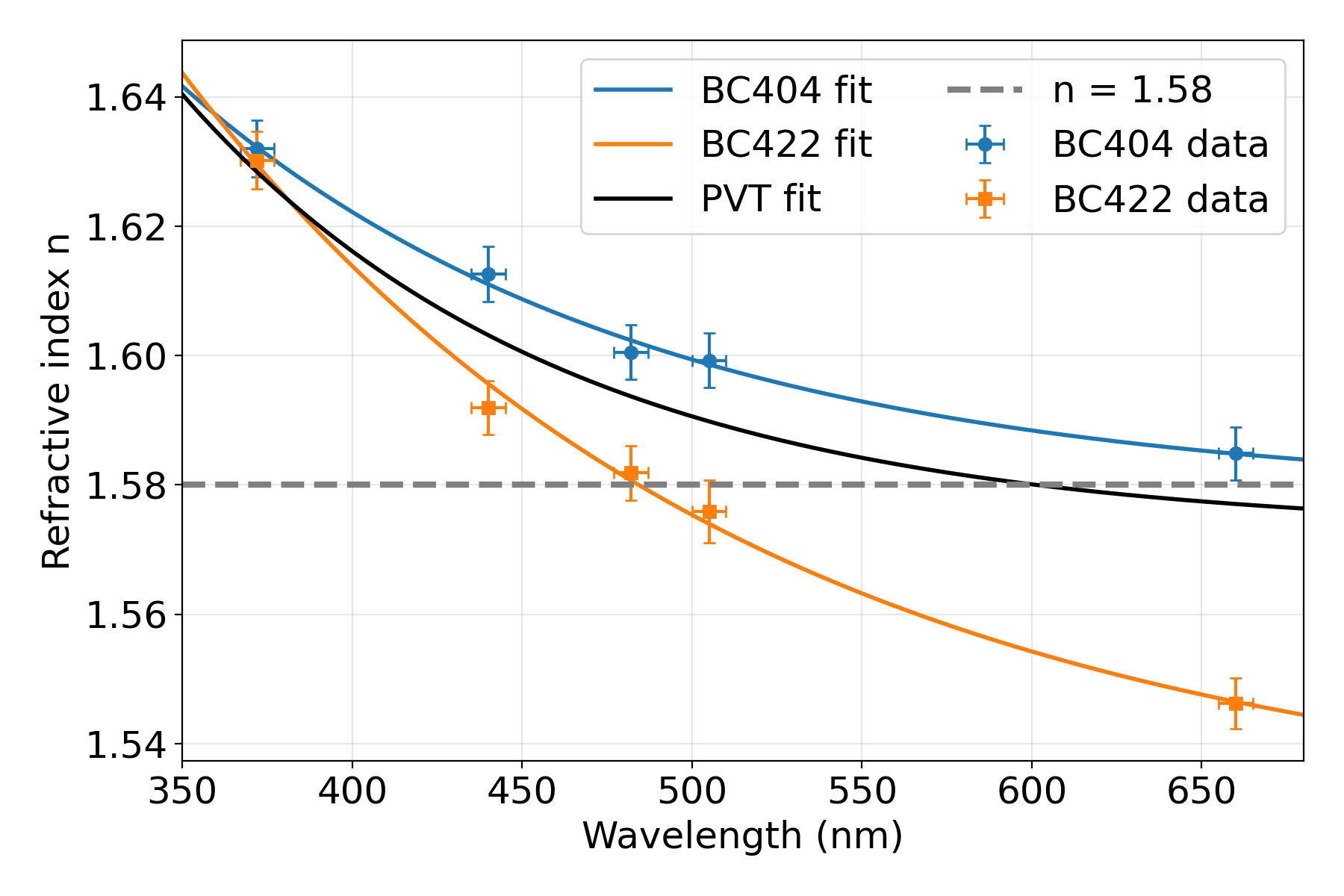} 
  \caption{Comparison of measured refractive indices for BC-404 and BC-422 to literature pure PVT and the manufacturer specification (n = 1.58). The manufacturer value fails to capture wavelength dependence, while literature PVT reproduces the general dispersion shape but differs in magnitude and curvature.}
  \label{fig:comparison}
\end{figure}

Figure~\ref{fig:comparison} compares our measured refractive indices for BC-404 and BC-422 with the manufacturer specification and literature values for pure PVT \cite{nakamura2015optical}. The pure-PVT data reproduce the general dispersion trend but do not capture all formulation-specific differences. Fluorescent additives can influence the optical response \cite{HamelPlasticScintillators}, although their individual contributions were not determined in this study.

The exponential model in Equation~\ref{eqn:expntl} provides excellent fits to our data across the measured wavelength range, with typical uncertainties in the weighted mean refractive index of 0.0055--0.0075. These uncertainties, dominated by systematic contributions from the rotation stage resolution ($0.1^{\circ}$) and micrometer positioning (0.18~mm), are sufficiently small to enable meaningful corrections in Monte Carlo simulations. Measurements at positive and negative rotation angles were consistent within their uncertainties, providing a check for sign-dependent angular bias.

The importance of refractive-index dispersion for photon transport depends on its overlap with the scintillation emission spectrum. At longer visible wavelengths, several formulations have fitted refractive indices below the manufacturer-specified value of 1.58. BC-430's red-shifted emission samples a different portion of the dispersion curve than the shorter-wavelength emission of the other formulations.

\subsection{Refractive-Index Effects on Simulated Response with Scintillator Bulk Absorption Omitted}

\begin{figure}[ht]
  \centering
  \includegraphics[width=0.45\textwidth]{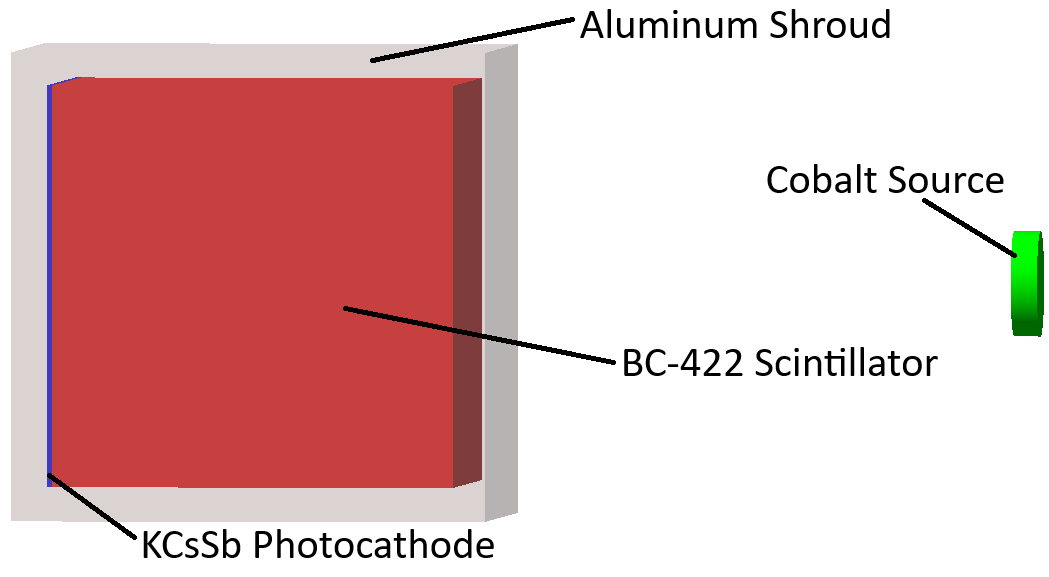} 
  \caption{Geant4 detector geometry. A BC-422 scintillator has a 0.1 cm thick KCsSb photocathode attached to count optical photons. The detector is surrounded by a 0.64 cm thick aluminum shroud for photon reflection. A cobalt puck of radius 1 cm, thickness 0.25 cm generates incident gammas.}
  \label{fig:GeantVis}
\end{figure}

To assess sensitivity to the refractive-index input, we perform Geant4 \cite{Geant4Software} (Version 11.4) simulations using a common detector geometry for the six scintillator formulations. The geometry consists of a $ 7.62 \times 7.62 \times 7.62~\mathrm{cm}^3 $ scintillator coupled to a modeled KCsSb photocathode \cite{Motta2005KCsSb} and surrounded by an aluminum shroud, as shown in Figure~\ref{fig:GeantVis}. Incident gamma rays have energies of 1.173 and 1.332~MeV, corresponding to the principal emissions of $ ^{60}\mathrm{Co} $. The scintillation emission spectra used for the individual materials are taken from the Saint-Gobain data sheets \cite{bc40040408_datasheet,bc418422_datasheet,bc430_datasheet}. The scintillator and KCsSb volumes were modeled in direct
contact, with no intervening air gap, window, or optical grease. The interface was specified as a \texttt{dielectric\_metal} optical surface using the UNIFIED model and a polished finish. BC-422 is used for the detailed depth-distribution comparison.

We implement the measured refractive-index dispersion as a material property table covering 370--660~nm. Optical surfaces at the scintillator-aluminum interface were defined with measured reflectivity and detection efficiency, while the scintillator-KCsSb boundary relied on Fresnel transmission and reflection calculated from wavelength-dependent refractive indices of both materials \cite{Emirhan2025surface, Motta2005KCsSb}. We did not include wavelength-dependent absorption in these initial simulations, as our absorption length measurements remain incomplete for most materials at visible wavelengths, as discussed in Section 3.3.

Table~\ref{tab:absorption_counts_comparison} compares the numbers of energy-deposition events exceeding 1.9~eV in the modeled KCsSb layer for the constant and measured refractive-index models. The measured-index model produces fewer threshold events for all six formulations, with reductions ranging from 1.2\% to 7.2\%. Changing the scintillator refractive index modifies the calculated Fresnel transmission and reflection at the scintillator--KCsSb interface. The comparison demonstrates sensitivity to the refractive-index input but does not separately quantify the contributions of individual interfaces.

\begin{table}[h]
\centering
\caption{Numbers of energy-deposition events exceeding 1.9~eV in the modeled KCsSb layer for the constant and measured refractive-index models. Percentage decreases are relative to the constant-index model. Each material and refractive-index configuration was simulated using 10,000 primary gamma rays, with energies selected with equal probability from the two energies 1.173 and 1.332 MeV}
\begin{tabular}{lrrr}
\hline
Material & $ n=1.58 $ & Measured $ n(\lambda) $ & Decrease (\%) \\
\hline
BC-400 & 7,926,564 & 7,580,980 & 4.4\% \\
BC-404 & 8,737,723 & 8,338,473 & 4.6\% \\
BC-408 & 8,159,070 & 7,751,799 & 5.0\% \\
BC-418 & 7,947,335 & 7,465,280 & 6.1\% \\
BC-422 & 3,117,873 & 2,893,720 & 7.2\% \\
BC-430 & 1,843,629 & 1,821,628 & 1.2\% \\
\hline
\end{tabular}
\label{tab:absorption_counts_comparison}
\end{table}

\begin{figure}[h]
  \centering
  \includegraphics[width=0.48\textwidth]{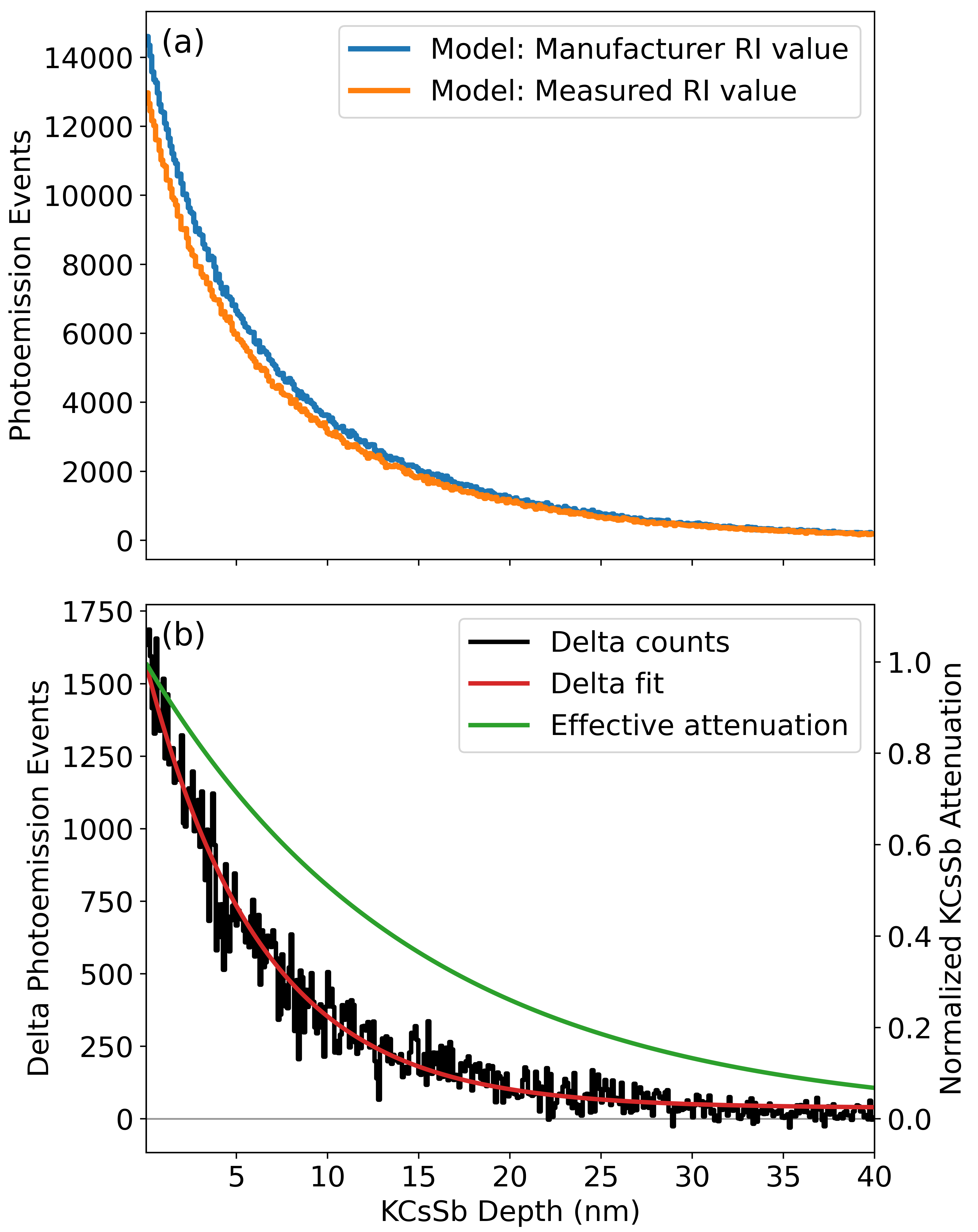} 
  \caption{Geant4 simulation of 10,000 gammas incident on the detector, uniformly sampled from $^{60}$Co photopeaks at 1.17 and 1.33~MeV. Results compare scintillator refractive index models, manufacturer specification, $n = 1.58$, and measured values, $n(\lambda) = 1.529 + 0.956e^{-0.00607\lambda}$. (a) Energy-deposition threshold events vs. KCsSb depth for each model. (b) Difference in events between models vs. the intensity-weighted harmonic mean absorption length of KCsSb.}
  \label{fig:comp}
\end{figure} 

Figure~\ref{fig:comp}a shows the energy-deposition threshold events for BC-422 as a function of depth into the modeled KCsSb layer. The measured refractive index model produces fewer threshold events overall, although the reduction is not uniform. Figure~\ref{fig:comp}b shows that this difference in counts does not track with the decay of the intensity-weighted harmonic mean absorption length of KCsSb. The disparity in the counts shows an overestimate in the optical photon population deeper in the photocathode material using the constant refractive index value.  This disparity indicates that ejected electrons, i.e. photoelectrons, from the surface of the photocathode material and its photoelectron escape depth (\textasciitilde{}10-30 nm for KCsSb) would be fewer than modeled with a constant index. The reduced number of ejected electrons would affect the modeled efficiency. Using the full range of the photoelectron escape depth for ejected electrons, the modeled efficiency is affected by 7.2\%. 

Photon interactions at the scintillator–aluminum boundary were included in both configurations. The reported energy-deposition threshold event differences reflect transport through the modeled optical system, including aluminum-boundary interactions and photocathode transmission. The contributions of the individual interfaces were not separately isolated. Including wavelength-dependent scintillator absorption would introduce additional losses that depend on wavelength and optical path length. For BC-430, measured absorption lengths range from 2.2 to 12.4~mm at four wavelengths between 372 and 505~nm, while the absorption length at 660~nm exceeds 830~mm. These differences indicate wavelength-dependent attenuation, but its effect on the simulated light yield and detected spectrum was not evaluated here.

The count differences demonstrate that the refractive-index input can affect simulated light-collection estimates. Their magnitude depends on the detector geometry, surface treatment, optical coupling, scintillation properties, and photosensor response; the present comparison does not establish refractive index as the dominant source of simulation uncertainty. Dispersion can also affect photon group velocities and arrival-time distributions. Its timing contribution was not quantified here and must be considered alongside variations in optical path length and scintillation emission time.

\subsection{Absorption Length Measurements}

Our variable-thickness transmission measurements successfully determined bulk absorption lengths for several wavelength-material combinations, particularly at 372~nm where all six scintillators exhibited measurable attenuation. At this UV wavelength, absorption lengths ranged from 2.2~$\pm$~0.1~mm for BC-430 to 37.7~$\pm$~2.7~mm for BC-422, with typical uncertainties of 10--15\%. These values reflect the increased absorption at higher photon energies, a general trend expected from electronic transitions in the PVT matrix and dopant molecules.

The most complete absorption characterization was obtained for BC-430, with finite absorption-length estimates at 372, 439, 482, and 505~nm. Its stronger absorption at these wavelengths is consistent with the role of fluorescent additives in shifting absorbed light toward longer-wavelength emission \cite{HamelPlasticScintillators}. The measurements do not identify the individual absorbing species or isolate their contributions. In contrast, BC-400, BC-404, and BC-408 have absorption-length lower bounds exceeding 750~mm at the longer measured wavelengths.

For these weakly-absorbing materials, our measurements provide only lower bounds on $L_{\mathrm{abs}}(\lambda)$. The limitation arises when the exponential attenuation over our maximum thickness of 40~mm for most materials and 100~mm for BC-400/404/408 becomes comparable to or smaller than our measurement uncertainty in integrated pulse area, typically 2--4\%. Under these conditions, the fit parameters $T_0(\lambda)$ and $L_{\mathrm{abs}}(\lambda)$ become poorly constrained, and the exponential model cannot distinguish between weak absorption and systematic variations in surface transmission. We report lower bounds using the adopted $ \Delta\chi^2=2.71 $ criterion.

The fitted normalization $ T_0(\lambda) $ in Equation~\ref{eqn:bulkattn} includes the incident light level, detector response, and thickness-independent transmission factors, including interface losses. These constant factors are separated from the thickness-dependent attenuation used to estimate $ L_{\mathrm{abs}}(\lambda) $. Differences in surface transmission between samples remain a potential source of uncertainty.

Extending sensitivity to longer absorption lengths would require a larger sample-thickness span, improved transmission precision, or both. Reference-beam monitoring or a dual-beam spectrophotometer could improve intensity stability and provide finer wavelength sampling. The attainable absorption-length precision would still depend on the transmitted-signal uncertainty and control of sample-to-sample surface differences.

Attempts to extend the measurements beyond 660~nm were limited by the signal obtained with the present PMT. Further wavelength coverage would require detectors with suitable spectral response at the wavelengths of interest. The present data do not establish the refractive-index or absorption behavior outside the measured wavelength interval.

\section{Summary and Conclusions}
We have characterized the wavelength-dependent refractive index and bulk absorption length of six commercial PVT scintillators: BC-400, BC-404, BC-408, BC-418, BC-422, and BC-430, at five wavelengths between 372 and 660~nm. The refractive-index measurements show material-specific dispersion that is not represented by the manufacturer-specified constant value of $ n=1.58 $. The empirical dispersion coefficients provided in Table~\ref{tab:exp_fit_coeffs} support optical-transport modeling within the measured wavelength interval.

Finite absorption-length estimates were obtained for all six materials at 372~nm, ranging from 2.2 to 37.7~mm. For BC-430, absorption lengths of 2.2--12.4~mm were determined at four wavelengths between 372 and 505~nm. For the remaining material and wavelength combinations, the measurements established lower bounds exceeding 750~mm. These estimates and bounds are reported in Table~\ref{tab:absorptionlngths}.

Geant4 comparisons with scintillator bulk absorption omitted demonstrate sensitivity to the refractive-index input. In the modeled geometry, replacing the constant refractive index with the measured dispersion reduced the number of energy-deposition events above 1.9~eV in the KCsSb layer by 1.2--7.2\%, with the largest reduction observed for BC-422. These differences describe the simulated threshold-event tally and should not be interpreted directly as equivalent changes in collected-photoelectron yield. Their magnitude depends on the detector geometry and optical configuration.

The reported measurements and fitted dispersion parameters provide formulation-specific inputs and constraints for scintillator detector simulations. Future measurements could extend wavelength coverage and improve sensitivity to longer absorption lengths through increased sample thickness and improved transmission precision.

\section*{Acknowledgements}
This manuscript has been authored by Lawrence Livermore National Security, LLC under Contract No. DE-AC52-07NA27344 with the U.S. Department of Energy. The United States Government retains, and the publisher, by accepting the article for publication, acknowledges that the United States Government retains a non-exclusive, paid-up, irrevocable, world-wide license to publish or reproduce the published form of this manuscript, or allow others to do so, for United States Government purposes.



\bibliographystyle{unsrtnat}
\bibliography{example}






\end{document}